\documentclass{rspublic}
\usepackage{epsf}   

\begin{document}

\title[Neutrino astronomy and gamma-ray bursts]{Neutrino astronomy and $\gamma$-ray bursts\footnote{
Review, R. Soc. Discussion Meeting on GRBs (To appear in Phil. Trans. R. Soc. Lond. A)}}

\author[E. Waxman]{Eli Waxman}

\affiliation{Physics Faculty, Weizmann Inst. of Science, Rehovot 76100, Israel}

\label{firstpage}

\maketitle
\begin{abstract}{gamma ray bursts, neutrinos, cosmic rays}

The construction of large volume detectors of high energy, $>1$~TeV, neutrinos is mainly driven by the search for extra-Galactic neutrino sources. The existence of such sources is implied by observations of ultra-high energy, $\ge10^{19}$~eV, cosmic-rays, the origin of which is a mystery. In this lecture I briefly discuss the expected extra-Galactic neutrino signal and the current state of the experimental efforts. Neutrino emission from $\gamma-$ray bursts (GRBs), which are likely sources of both high energy protons and neutrinos, is discussed in some detail. The detection of the predicted GRB neutrino signal, which may become possible in the coming few years, will allow one to identify the sources of ultra-high energy cosmic-rays and to resolve open questions related to the underlying physics of GRB models. Moreover, detection of GRB neutrinos will allow one to test for neutrino properties (e.g. flavor oscillations and coupling to gravity) with an accuracy many orders of magnitude better than is currently possible.

\end{abstract}

\section{Introduction}
\label{sec:intro}

The detection of MeV neutrinos from the Sun enabled direct observations of nuclear reactions in the core of the Sun, as well as studies of fundamental neutrino properties (Bahcall 2005). Existing MeV neutrino "telescopes" are also capable of detecting neutrinos from supernova explosions in our local Galactic neighborhood, at distances $<100$~kpc, such as supernova 1987A. The detection of neutrinos emitted by SN1987A provided a direct observation of the core collapse process and constraints on neutrino properties (Raffelt 2005). 

MeV neutrino detectors are unlikely to enable the detection of an extra-Galactic neutrino signal. In order to extend the distance accessible to neutrino astronomy to extra-Galactic scales, high energy, $>1$~TeV, neutrino telescopes are being constructed (Halzen 2005). The advantage in going to high energy is twofold: The neutrino interaction cross section increases with energy, and the construction of larger detectors is possible at higher energy. 

The target sources of high energy neutrino telescopes are different than those of low energy neutrino telescopes, since, unlike MeV neutrinos, TeV neutrinos are not produced in thermonuclear interactions in stars. $>1$~TeV neutrino telescopes are aiming at detecting "cosmic accelerators," which produce high energy particles
(Waxman 2005). The existence of extra-Galactic high-energy neutrino sources is implied by the existence of cosmic-rays with energies extending to $>10^{20}$~eV: The $>10^{19}$~eV cosmic-ray flux is likely dominated by extra-Galactic sources of protons (Nagano \& Watson 2000) (see, however, Watson 2004), and the interaction of such high energy protons with radiation fields or gas (either near or far from the source) leads to the production of charged pions which decay to produce neutrinos (e.g. $p+\gamma\rightarrow n+\pi^+$, $\pi^+\rightarrow\mu^++\nu_\mu\rightarrow e^++\nu_\mu+\bar{\nu}_\mu+\nu_e$). 

The origin of the highest energy, $>10^{19}$~eV, cosmic rays (UHECRs) is a mystery: The identity of the UHECR sources and the mechanism(s) by which particles are accelerated to such extreme energies are unknown. Identifying the sources and understanding the acceleration mechanism are among the most interesting and important challenges of high energy astrophysics (Bhattacharjee \& Sigl 2000, Waxman 2004a). Identifying the UHECR sources and probing the physical mechanisms driving them are among the main science targets of high energy neutrino detectors (Waxman 2005). These issues would not be discussed, however, in detail in this review, which is focused on the relation between neutrino astronomy and gamma-ray bursts (GRBs).

A phenomenological, model independent discussion of extra-Galactic high energy neutrino sources is presented in \S~\ref{sec:WB}. It is shown that the UHECR observations set an upper bound to the extra-Galactic high energy neutrino flux. The experimental implications of this bound and the current state of the experimental effort are briefly reviewed in \S~\ref{sec:nu_tel}. High energy neutrino production in GRBs is discussed in \S~\ref{sec:GRBnu}. Testing for fundamental neutrino properties using GRB neutrinos is discussed in \S~\ref{sec:nu_phys}. A summary and a brief discussion of future prospects is given in \S~\ref{sec:discussion}.

\section{Cosmic accelerators and the neutrino background}
\label{sec:WB}

The cosmic-ray spectrum extends to energies $>10^{20}$~eV. The flux at the highest observed energies is rather low,
\begin{equation}\label{eq:jCR}
    j_{CR}(>10^{20}{\rm eV})\approx 1/{\rm km^2}/100{\rm yr}/(2\pi\,{\rm sr}).
\end{equation}
The flattening of the spectrum at $\sim10^{19}$~eV and the isotropic distribution of arrival directions suggest that the cosmic-ray flux above $\sim10^{19}$~eV is dominated by extra-Galactic sources (Nagano \& Watson 2000). There is some evidence that the highest energy particles are light nucleons (presumably protons), but the identity of the cosmic-ray primaries is still debated (e.g. Watson 2004). Under the assumption that the UHECRs are extra-Galactic protons, the observed flux and spectrum of UHECRs determines the rate (per unit time and volume) and the spectrum with which high energy protons are produced by extra-Galactic sources to be (Waxman 1995b, Bahcall \& Waxman 2003)
\begin{equation}
    E_p^2\frac{d\dot{N}_p}{dE_p}=0.65\times 10^{44} {\rm erg~Mpc^{-3}~yr^{-1}}\phi(z). 
\label{eq:energyrate}
\end{equation}
Here, $\phi(z)$ accounts for redshift evolution and $\phi(z=0)=1$. The flux and spectrum above $10^{19}$~eV are only weakly dependent on $\phi(z)$ since proton energy loss limits the protons' propagation distance. 
Protons of energy exceeding the threshold for pion production in interaction with the cosmic microwave background photons, $\sim5\times10^{19}$~eV, lose most of their energy over a time short compared to the age of the universe, or, equivalently, over a propagation distance $d_{\rm GZK}\simeq100$~Mpc, which is short compared to the Hubble distance (the "GZK effect"; Greisen 1966, Zatsepin \& Kuzmin 1966). 

The identity of the UHECR sources as well as the acceleration mechanism are not yet known. The stringent constraints, which are imposed on the properties of possible UHECR sources by the high energies observed, rule out almost all source candidates, and suggest that GRBs and active galactic nuclei (AGN) are the most plausible sources (Bhattacharjee \& Sigl 2000, Waxman 2004a). Active galactic nuclei are disfavored as sources of UHECRs, since no AGN powerful enough to allow acceleration to $>10^{20}$~eV is present out to a distance $\sim d_{\rm GZK}$ from earth. The suggested association between GRB and UHECR sources is based on two arguments (Waxman 1995a, Waxman 2004b): (i) The constraints that UHECR sources must satisfy to allow proton acceleration to $>10^{20}$~eV are similar to those inferred for GRB sources from $\gamma$-ray observations, and (ii) The average energy generation rate of UHECR's is similar to the $\gamma$-ray generation rate of GRBs (Associations based on different arguments were suggested in Milgrom \& Usov 1995 and in Vietri 1995). 

Adopting the hypothesis that UHECRs are extra-Galactic protons, an upper bound to the extra-Galactic high energy neutrino intensity may be obtained by the following argument. Let us consider first the neutrino intensity that would have been produced had the high energy cosmic-ray protons lost all their energy to pion production in $p\gamma$ or $pp(n)$ interactions. In this case, the resulting intensity of high energy muon neutrinos (including muon and anti-muon neutrinos, neglecting mixing) would have been (Waxman \& Bahcall 1999, Bahcall \& Waxman 2001)
\begin{equation}
E_\nu^2\Phi_\nu=E_\nu^2\Phi_\nu^{WB}=
\frac{1}{4}\frac{c}{4\pi}E_p^2\left(\frac{d\dot{N}_p}{dE_p}\right)_{z=0}\xi_z t_H.
\label{eq:WB}
\end{equation}
Here $t_H$ is the Hubble time and the factor of $1/4$ accounts for the fact that $\approx1/4$ of the pions' energy is converted to muon neutrino energy (since charged pions and neutral pions are produced with roughly equal probability, and $\approx1/2$ of the charged pion energy is channelled into muon neutrinos). $\xi_z$ is (a dimensionless parameter) of order unity, which depends on the redshift evolution of $E_p^2d\dot{N}_p/dE_p$ (see eq.~\ref{eq:energyrate}). $\Phi_\nu^{WB}$ is an upper bound to the neutrino intensity produced by sources which, like GRBs and AGN jets, are optically thin for high-energy nucleons to $p\gamma$ and $pp(n)$ interactions (since in such sources protons lose only a small fraction of their energy to pion production before escaping the source). For sources of this type, therefore,
\begin{equation}
\label{eq:WB_bound}    
E_\nu^2\Phi_\nu<E_\nu^2\Phi_\nu^{WB}
=2\times10^{-8}\xi_z\left[\frac{(E_p^2d\dot{N}_p/dE_p)_{z=0}}{10^{44}{\rm erg/Mpc^3yr}}\right]
{\rm GeV\,cm}^{-2}{\rm s}^{-1}{\rm sr}^{-1}.
\end{equation}
In order to obtain a conservative upper bound, we adopt $(E_p^2d\dot{N}_p/dE_p)_{z=0}=10^{44}{\rm erg/Mpc^3yr}$ and a rapid redshift evolution, $\phi(z)=(1+z)^3$ up to $z=2$, following the evolution of star formation rate. This evolution yields $\xi_z\approx3$.

\section{Neutrino telescopes}
\label{sec:nu_tel}

\begin{figure}[htbp]
\epsfxsize=12cm
\centerline{\epsfbox{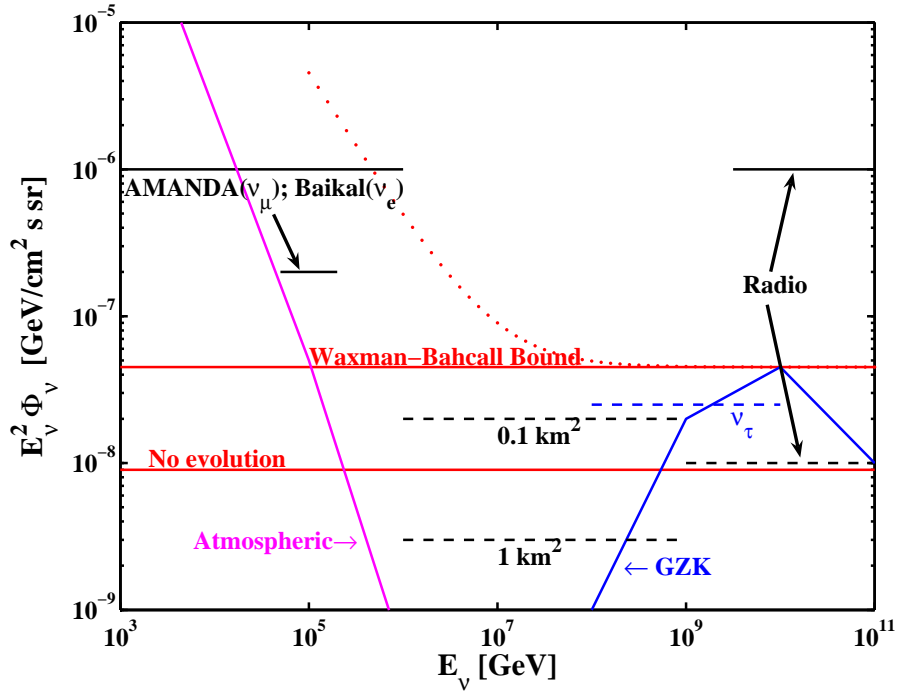}}
\caption{The upper bound imposed by UHECR observations on the extra-Galactic high energy muon neutrino intensity (lower-curve: no evolution of the energy production rate, upper curve: assuming evolution following star formation rate), compared with the atmospheric neutrino background and with the experimental upper bounds (solid lines) of optical Cerenkov experiments, BAIKAL (Balkanov et al. 2002) and AMANDA (Ahrens et al. 2003; Halzen 2005), and of coherent Cerenkov radio experiments (RICE, Kravchenko et al. 2003; GLUE, Gorham et al. 2004, Silvestri et al. 2004). The dotted curve is the maximum contribution due to possible extra-Galactic component of lower-energy, $<10^{17}$~eV, protons, that may be "hidden under"  the Galactic heavy nuclei flux (Waxman \& Bahcall 1999, Bahcall \& Waxman 2001). The curve labelled "GZK" shows the intensity due to interaction with micro-wave background photons. Dashed curves show the expected sensitivity of 0.1~Gton (AMANDA, ANTARES, NESTOR) and 1~Gton (IceCube, km3net) optical Cerenkov detectors (Halzen 2005), of the coherent radio Cerenkov (balloon) experiment ANITA (Silvestri et al. 2004) and of the Auger air-shower detector (sensitivity to $\nu_\tau$) (Saltzberg 2005). Space air-shower detectors (OWL-AIRWATCH) may also achieve the sensitivity required to detect fluxes lower than the WB bound at energies $>10^{18}$~eV (Saltzberg 2005).}
\label{fig:WBbound}
\end{figure}

The UHECR upper bound on the extra-Galactic neutrino intensity, eq.~(\ref{eq:WB_bound}), is compared in fig.~\ref{fig:WBbound} with current experimental limits, and with the expected sensitivity of planned neutrino telescopes. The figure indicates that km-scale (i.e. giga-ton) neutrino telescopes are needed to detect the expected extra-Galactic flux in the energy range of $\sim1$~TeV to $\sim1$~PeV. Much larger effective volume is required to detect the flux at higher energy. 

The AMANDA detector operating in the south pole has already achieved an effective detector mass of $\sim0.1$~Gton. It detects neutrinos mainly by measuring the optical Cerenkov light emitted by high energy muons produced by muon neutrino interactions in the ice. The IceCube extension of AMANDA, planned to reach a $\sim1$~Gton effective detector mass, is well under way: The AMANDA detector consists of 660 optical modules, and during the 05/06 (south pole) summer season alone 600 additional IceCube optical modules were successfully deployed. The full IceCube detector, consisting of 4800 modules, is expected to be completed in 4~yrs. A $\sim0.1$~Gton optical Cerenkov experiment, ANTARES, is currently under construction in the Mediterranean sea.  The ANTARES experiment has been successfully operating a single line of 75 optical modules since March 06, and is fully funded to be expanded in the coming few years to reach a $\sim0.1$~Gton detector mass. In parallel, R\&D has begun on a $\sim1$~Gton Mediterranean optical Cerenkov detector within the km3net collaboration. For more details on these optical Cerenkov neutrino telescopes see Halzen (2005) and references therein, and http://icecube.wisc.edu/, http://antares.in2p3.fr/, http://www.nestor.org.gr/hena/intro/intro.htm, http://www.km3net.org/.

At ultra high energies, $\gg1$~PeV, much larger detectors, $\gg1$~Gton, are required. The construction of sufficiently large detectors may be possible by taking advantage of the large coherent Cerenkov radio signals emitted by the particle showers produced by the interaction of ultra high energy neutrinos in the ice. The RICE detector (Kravchenko et al. 2003) is aiming at detecting these radio signals using antennas buried in the ice, while the ANITA experiment (Silvestri et al. 2004) will be using balloon born antennas over the south pole (a full engineering flight is planned for ANITA at the end of 06). The large effective mass required at ultra high energies may also be achieved by large cosmic ray detectors (such as Auger) searching for neutrino induced air showers. For discussion of these experimental techniques see Saltzberg (2005) and references therein, and http://www.ps.uci.edu/~anita/, http://www.bartol.udel.edu/~spiczak/rice/rice.html.

%\section{The neutrino background from star-burst galaxies}
%\label{sec:starburst}

\section{GRB neutrinos}
\label{sec:GRBnu}

Gamma-ray bursts are short, typically tens of seconds long, flashes of gamma-rays, carrying most of their energy in $>1$~MeV photons. The detection in the past few years of "afterglows," delayed X-ray, optical and radio emission from GRB sources, proved that the sources lie at cosmological distances, and provided strong support for the following scenario of GRB production (for recent reviews see Piran 2005, M\'esz\'aros 2006). The energy source is believed to be rapid mass accretion on a newly formed solar-mass black hole. Observations suggest that the formation of the central compact object is associated, at least in some cases, with type Ib/c supernovae. The energy release drives an ultra-relativistic, $\Gamma\sim10^{2.5}$, plasma outflow. The emission of $\gamma$-rays is assumed to be due to internal collisionless shocks within the relativistic wind, the "fireball," which occur at a large distance from the central black-hole due to variability in the wind emitted from the central "engine." It is commonly assumed that electrons are accelerated to high energy within the collisionless shocks, and that synchrotron emission from these shock accelerated electrons produces the observed $\gamma$-rays. 

If protons are present in the wind, as assumed in the fireball model, they would also be accelerated to high energy in the region were electrons are accelerated. As mentioned in \S~\ref{sec:WB}, the high energy to which protons may be accelerated in GRB winds and the rate (per unit time and volume) at which GRBs produce energy make GRBs likely sources of UHECR protons (see Waxman 2001 for a pedagogical review). If protons are indeed accelerated to high energy in GRBs, then high energy neutrino emission is also expected.

\subsection{100 TeV fireball neutrinos}
\label{sec:fireball_nu}

Protons accelerated in the fireball internal shocks, where GRB $\gamma$-rays are expected to be produced, lose energy through photo-production of pions in interactions with fireball photons. The decay of charged pions produced in this interaction results in the production of high energy neutrinos. The key relation is between the observed photon energy, $E_\gamma$, and the accelerated proton's energy, $E_p$, at the threshold of the
$\Delta$-resonance. In the observer frame,
\begin{equation}
E_\gamma \,E_{p} = 0.2 \, {\rm GeV^2} \, \Gamma^2\,.
\label{eq:keyrelation}
\end{equation}
For $\Gamma\approx10^{2.5}$ and $E_\gamma=1$~MeV, we see that characteristic proton energies $\sim 10^{16}$~eV are required to produce pions. Since neutrinos produced by pion decay typically carry $5\%$ of the proton energy, production of $\sim 10^{14}$~eV neutrinos is expected (Waxman \& Bahcall 1997).

The fraction of energy lost by protons to pions, $f_\pi$, is $f_\pi\approx0.2$ (Waxman \& Bahcall 1997, Waxman 2001). Assuming that GRBs generate the observed UHECRs, the expected GRB muon and anti-muon neutrino flux may be estimated using eq.~(\ref{eq:WB}) (Waxman \& Bahcall 1997, 1999), 
\begin{equation}
E_\nu^2\Phi_{\nu}\approx 0.8\times10^{-8}{f_\pi\over0.2}{\rm GeV\,cm}^{-2}{\rm s}^{-1}{\rm
sr}^{-1}. \label{eq:JGRB}
\end{equation}
This neutrino spectrum extends to $\sim10^{16}$~eV, and is suppressed at higher energy due to energy loss of pions and muons (Waxman \& Bahcall 1997, 1999, Rachen \& P. M\'esz\'aros 1998; for the contribution of Kaon decay at high energy see Asano \& Nagataki 2006). Eq.~(\ref{eq:JGRB}) implies a detection rate of $\sim20$ neutrino-induced muon events per year (over $4\pi$~sr) in a cubic-km detector (Waxman \& Bahcall 1997, Alvarez-Mu\~{n}iz et al. 2000, Guetta et al. 2004). 

Since GRB neutrino events are correlated both in time and in direction with gamma-rays, their detection is practically background free. The main background is due to atmospheric neutrinos, which produce neutrino-induced muons, travelling in a direction lying within a cone of opening angle $\Delta\theta$ around some direction, at a rate
\begin{equation}\label{eq:atmo}
    J_{\nu\rightarrow\mu}^A\simeq4\times10^{-3}\left(\frac{\Delta\theta}{0.5^o}\right)^2
\left(\frac{E}{100\rm TeV}\right)^{-\beta}\,{\rm km^{-2}yr^{-1}},
\end{equation}
with $\beta=1.7$ for $E<100$~TeV and $\beta=2.5$ for $E>100$~TeV. At high energies, the neutrino induced muon propagates at nearly the same direction as the incoming neutrino, and km-scale neutrino telescopes will be able to determine the incoming neutrino direction to better than $\sim0.5^o$. For a known source direction, therefore, the neutrino search is practically background free.

\subsection{TeV neutrinos}
\label{sec:TeV}

The 100~TeV neutrinos discussed in the previous section are produced at the same region where GRB $\gamma$-rays are produced. Their production is a generic prediction of the fireball model. It is a direct consequence of the assumptions that energy is carried from the underlying engine as kinetic energy of protons and that $\gamma$-rays are produced by synchrotron emission of shock accelerated particles. Neutrinos may be produced also in other stages of fireball evolution, at energies different than 100~TeV. The production of these neutrinos is dependent on additional model assumptions. We discuss below some examples related to the GRB progenitor. For a discussion of neutrino emission during the afterglow phase see, e.g., Dai \& Lu (2001). For a more detailed discussion see M\'esz\'aros (2002) and references therein.

The most widely discussed progenitor scenarios for long-duration GRBs involve core collapse of massive stars. In these "collapsar" models, a relativistic jet breaks through the stellar envelope to produce a GRB. For extended or slowly rotating stars, the jet may be unable to break through the envelope. Both penetrating (GRB producing) and "choked" jets can produce a burst of $\sim5$~TeV neutrinos by interaction of accelerated protons with jet photons, while the jet propagates in the envelope (M\'esz\'aros \& E. Waxman 2001, Razzaque et al. 2004). The estimated event rates may exceed $\sim10^2$ events per yr in a km-scale detector, depending on the ratio of non-visible to visible fireballs. A clear detection of non-visible GRBs with neutrinos may be difficult due to the low energy resolution for muon-neutrino events, unless the associated supernova photons are detected. In the two-step "supranova" model, interaction of the GRB blast wave with the supernova shell can lead to detectable neutrino emission, either through nuclear collisions with the dense supernova shell or through interaction with the intense supernova and backscattered radiation field (Dermer \& Atoyan 2003, Guetta \& Granot 2003, Razzaque et al. 2003).

\subsection{GRB neutrinos and SWIFT}
\label{sec:swift}

Recent observations with SWIFT revealed new and unexpected GRB characteristics, which may require modifications and extensions of GRB models. Such extensions and modifications may have implications for the predicted neutrino signal. The detection of GRB neutrinos may therefore allow one to discriminate between different models suggested for explaining these unexpected GRB characteristics.  

Possibly the most intriguing new characteristic is the existence of strong X-ray flares following the prompt $\gamma$-ray emission with delays of hundreds of seconds to days, sometimes carrying energy comparable to that of the prompt emission (Burrows 2006). These flares may be explained as due to delayed internal shocks within the fireball wind, due to a prolonged activity of the "central engine" (e.g. Fan \& Wei 2005). If this interpretation is correct, these delayed flares should be accompanied by emission of PeV neutrinos, which may be detectable by km-scale neutrino telescopes (Murase \& Nagataki 2006).

\section{Neutrino physics and astrophysics}
\label{sec:nu_phys}

Detection of high energy neutrinos from GRBs will provide information on fundamental neutrino properties (Waxman \& Bahcall 1997). 

Detection of neutrinos from GRBs could be used to test the simultaneity of neutrino and photon arrival to an accuracy of $\sim1$~s. It is important to emphasize here that since the background level of
neutrino telescopes is very low, see eq.~(\ref{eq:atmo}), the detection of a single neutrino from the direction of a GRB on a time sale of months after the burst would imply an association of the neutrino with the burst and will therefore establish a time of flight delay measurement. Such a measurement will allow one to test for violations of Lorentz invariance (as expected due to quantum gravity effects) (Waxman \& Bahcall 1997, Amelino-Camelia et al. 1998, Colman \& Glashow 1999, Jacob \& Piran 2006), and to test the weak equivalence principle, according to which photons and neutrinos should suffer the same time delay as they pass through a gravitational potential. With $1{\rm\ s}$ accuracy, a burst at $1{\rm\ Gpc}$ would reveal a fractional difference in (photon and neutrino) speed of $10^{-17}$, and a fractional difference in gravitational time delay of order $10^{-6}$ (considering the Galactic potential alone). Previous applications of these ideas to supernova 1987A (see Bahcall 1989 for review), yielded much weaker upper limits: of order $10^{-8}$ and $10^{-2}$ respectively.

High energy neutrinos are expected to be produced in GRBs by the decay of charged pions, which lead to the production of neutrinos with flavor ratio $\Phi_{\nu_e}:\Phi_{\nu_\mu}:\Phi_{\nu_\tau}=1:2:0$ (here $\Phi_{\nu_l}$ stands for the combined flux of $\nu_l$ and $\bar\nu_l$). Neutrino oscillations then lead to an observed flux ratio on Earth of
$\Phi_{\nu_e}:\Phi_{\nu_\mu}:\Phi_{\nu_\tau}=1:1:1$ (Learned \& Pakvasa 1995). Up-going $\tau$'s, rather than $\mu$'s, would be a distinctive signature of such oscillations. It has furthermore been been pointed out that flavor measurements of astrophysical neutrinos may help determining the mixing parameters and mass hierarchy (Winter 2006), and may possibly enable one to probe new physics (Learned \& Pakvasa 1995,Athar et al. 2000). 

\begin{figure}[htbp]
\epsfxsize=14cm
\centerline{\epsfbox{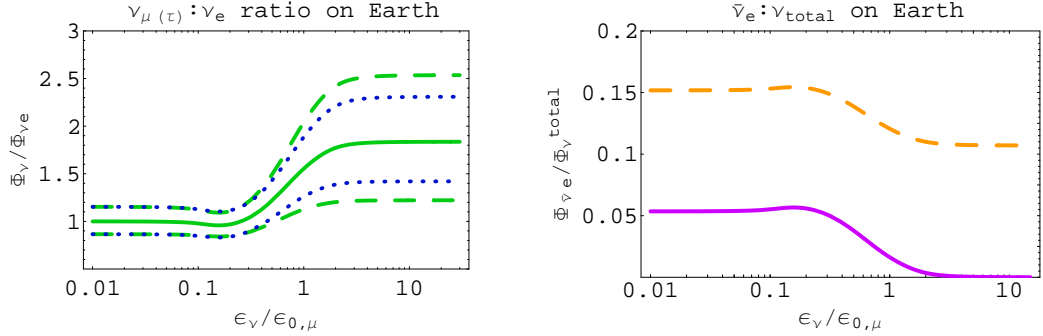}}
\caption{Flavor and anti-particle content of the flux of astrophysical neutrinos produced by pion decay, for pion energy spectrum (at production) $dn/dE_\pi\propto E_\pi^{-2}$ and electromagnetic pion energy loss rate $d E_\pi/dt\propto E_\pi^2$, as expected for internal shocks in GRB fireballs. Left: The ratio between $\Phi_{\nu_{\mu}(\nu_\tau)}$ and $\Phi_{\nu_e}$ (solid line), with 90\% CL lines of $\nu_\mu$ (dashed) and $\nu_\tau$ (dotted) fluxes (here, $\Phi_{\nu_l}$ stands for the combined flux of $\nu_l$ and $\bar\nu_l$). Right: The ratio of $\bar\nu_e$ to total $\nu$ flux on Earth, solid (dashed) line for neutrinos produced by  $p\gamma$ ($pp$) interactions. $E_{0,\mu}$ is the muon energy for which the muon life time is comparable to its electromagnetic energy loss time. $E_{0,\mu}\approx10^3$~TeV for internal shocks in GRB fireballs.}
\label{fig:fNu}
\end{figure}
It is interesting to point out here that flavor ratio measurements may provide interesting constraints not only on fundamental neutrino properties, but also on the astrophysical models. Although a $1:1:1$ flavor ratio appears to be a robust prediction of models where neutrinos are produced by pion decay, energy dependence of the flavor ratio is a generic feature of models of high energy astrophysical neutrino sources (Kashti \& Waxman 2005). Pions are typically produced in environments where they may suffer significant energy losses prior to decay, due to interaction with radiation and magnetic fields (Waxman \& Bahcall 1997, Rachen \& M\'esz\'aros 1998). Since the pion life time is shorter than the muon's, at sufficiently high energy the probability for pion decay prior to significant energy loss is higher than the corresponding probability for muon decay. This leads to suppression at high energy of the relative contribution of muon decay to the neutrino flux. The flavor ratio is modified to $0:1:0$ at the source, implying $1:1.8:1.8$ ratio on Earth (Kashti \& Waxman 2005). The energy at which the transition from $0:1:0$ to $1:1.8:1.8$ ratio takes place depends on the energy density (of magnetic fields and radiation) at the source. Detection of the flavor ratio transition will therefore provide constraints on the properties of the source. Figure~\ref{fig:fNu} presents the expected energy dependence of flavor and anti-particle content for internal shocks in GRB fireballs. Since the transition is expected at $\sim100$~TeV, it may be detectable by km-scale neutrino telescopes.

\section{Discussion}
\label{sec:discussion}

Several large volume high energy neutrino telescopes are currently operating and under construction (see \S~\ref{sec:nu_tel}): The optical Cerenkov detector AMANDA is operating at the south pole with an effective detector mass of 0.1~Gton for neutrinos in the 1~TeV--1~PeV energy range, and being expanded to a 1~Gton detector (IceCube); 0.1~Gton (for neutrinos in the 1~TeV--1~PeV energy range) optical Cerenkov detectors (ANTARES, NESTOR) are being constructed in the Mediterranean sea, and a 1~Gton Mediterranean detector is at an R\&D phase (km3NeT); Still larger detectors are being built for higher energy neutrinos, $\gg1$~PeV, utilizing the coherent radio Cerenkov emission of neutrino induced showers in the ice (ANITA, RICE). These detectors will reach in the coming few years a sensitivity which may allow the detection of extra-Galactic neutrino signals (see \S~\ref{sec:WB} and \S~\ref{sec:nu_tel}).

The detection of extra-Galactic neutrinos will allow us to answer some of the most interesting and important questions of high energy astrophysics (see \S~\ref{sec:WB}): It may allow us to identify the sources of UHECRs, to identify the UHECR primaries, and to probe the physics of particle acceleration to ultra-high energy. 

GRBs are likely sources of UHECRs and high energy neutrinos (see \S~\ref{sec:GRBnu}). The production of 100~TeV neutrinos in the region where GRB $\gamma$-rays are produced is a generic prediction of the fireball model (see \S~4a). It is a direct consequence of the {\it assumptions} that energy is carried from the underlying engine, most likely a (few) solar mass black hole, as kinetic energy of protons and that $\gamma$-rays are produced by synchrotron emission of shock accelerated particles. The detection of the predicted neutrino signal will therefore provide strong support for the validity of underlying model assumptions, which is difficult to obtain using photon observations due to the high optical depth in the vicinity of the GRB "engine." For example, an alternative model where the energy is carried away from the source as pure Poynting flux and where $\gamma$-rays are produced by magnetic field dissipation (e.g. Lyutikov et al. 2003) is unlikely to be ruled out by photon observations, but will be strongly disfavored by the detection of 100~TeV neutrinos. The predicted neutrino intensity implies a detection of $\sim20$ muon induced neutrino events per yr in a km-scale neutrino detector. Since these events should be correlated in direction with GRB $\gamma$-rays, the search for GRB neutrinos is essentially background free (see \S~4a). 

Recent observations with SWIFT revealed new and unexpected GRB characteristics, which may require modifications and extensions of GRB models. Such extensions and modifications may have implications for the predicted neutrino signal and the detection of GRB neutrinos may allow one to discriminate between different models. In particular, if the delayed X-ray flares are due to delayed internal shocks within the fireball wind, due to a prolonged activity of the "central engine," emission of PeV neutrinos should accompany the flares and may be detectable by Gton neutrino telescopes (see \S~4c). Neutrino observations may also allow one to probe GRB progenitors, as lower energy, $\sim1$~TeV, neutrinos are expected to be produced if GRBs are associated with the collapse of massive stars (see \S~4b).

The discussion of GRB neutrino emission demonstrates that in addition to identifying the sources of UHECRs, high-energy neutrino telescopes can provide a unique probe of the physics of these sources. In addition, detection of high energy neutrinos from GRB sources will provide information on fundamental neutrino properties (see \S~\ref{sec:nu_phys}). Detection of GRB neutrinos could be used to test the simultaneity of neutrino and photon arrival to an accuracy of $\sim1$~s. This will allow one to test for violations of Lorentz invariance and to test the weak equivalence principle, according to which photons and neutrinos should suffer the same time delay as they pass through a gravitational potential. With $1{\rm\ s}$ accuracy, a burst at $1{\rm\ Gpc}$ would reveal a fractional difference in (photon and neutrino) speed of $10^{-17}$, and a fractional difference in gravitational time delay of order $10^{-6}$ (considering the Galactic potential alone). Previous applications of these ideas to supernova 1987A, yielded much weaker upper limits: of order $10^{-8}$ and $10^{-2}$ respectively.

The decay of charged pions in GRB fireballs is expected to produce electron and muon neutrinos only. Neutrino oscillations would then lead to an observed flux ratio on Earth of $\Phi_{\nu_e}:\Phi_{\nu_\mu}:\Phi_{\nu_\tau}=1:1:1$. Up-going $\tau$'s, rather than $\mu$'s, would be a  distinctive signature of such oscillations. Flavor measurements of astrophysical neutrinos may furthermore help determining the mixing parameters and mass hierarchy, and may possibly enable one to probe new physics.

Finally, it is interesting to point out that flavor ratio measurements may provide interesting constraints not only on fundamental neutrino properties, but also on the astrophysical models. Pion and muon energy losses are expected to modify the flavor ratio observed on Earth to $1:1.8:1.8$ at high energies (see \S~\ref{sec:nu_phys}). The energy at which the transition from $0:1:0$ to $1:1.8:1.8$ ratio takes place depends on the energy density (of magnetic fields and radiation) at the source and therefore provides constraints on the properties of the source. For GRB sources the transition is expected at $\sim100$~TeV, and may be detectable by km-scale neutrino telescopes.

\end{document}